\documentclass[prd,twocolumn,aps,amsmath,amssymb,showpacs,10pt]{revtex4-1}

\usepackage{amsbsy}
\usepackage{graphicx}
\usepackage{bm}
\usepackage{graphics}
\usepackage{latexsym}
\usepackage[active]{srcltx}
\usepackage{amsbsy}
\usepackage{latexsym,amsfonts,amsthm,amsmath,amscd,amssymb,color}
\usepackage{graphicx}
\usepackage{amsbsy}
\newcommand{\sect}[1]{\setcounter{equation}{0}\section{#1}}

\newcommand{\myatop}[2]{\genfrac{}{}{0pt}{}{#1}{#2}}
\def\spazio#1{\vrule height#1em width0em depth#1em}

\def\spazio#1{\vrule height#1em width0em depth#1em}

\begin{document}

\begin{abstract}
This paper is a prosecution of a previous work where we presented a unified two-fermion
covariant scheme which produced very precise results for the masses of light and heavy mesons.
We extend the analysis to some radiative decays of mesons
$\Upsilon,\,\chi_{b2}\,,h_b\,,\chi_{b1}\,,\chi_{b0}\,,\eta_b\,$ 
and we calculate their branching ratios and their widths.  For most of them 
we can make a comparison with experimental data finding a good agreement.For the decays
for which data are not available we compare ours with other recent theoretical previsions.
\medskip


\end{abstract}

\title{\bf Relativistic two-body calculation of $b\bar{b}\,$-mesons radiative decays.}

\author{A. Barducci}
\affiliation{Dipartimento di Fisica, Universit\`a di Firenze,
Italy} \affiliation{Istituto Nazionale di Fisica Nucleare, Sezione
di Firenze} 
\author{R. Giachetti}
\affiliation{Dipartimento di Fisica, Universit\`a di Firenze,
Italy} \affiliation{Istituto Nazionale di Fisica Nucleare, Sezione
di Firenze} 
\author{E. Sorace}
\affiliation{Istituto Nazionale di Fisica Nucleare, Sezione di
Firenze}

\maketitle

%
%
%
%

%
%
%
\sect{Introduction} \label{Sec_introduction}
%
%
%
Potential models have long been used to  investigate the meson spectrum 
\cite{EGKKLY,LSGI,PotMod,RR-BBD,Gro}. After the pioneering non relativistic work \cite{EGKKLY}, 
it emerged with evidence that the relativity effects are important for the description of
mesons \cite{Rich}, both light and heavy, so that the following papers have eventually included
relativistic corrections, either by perturbation methods or by covariant approaches
in order to reach a reasonable precision \cite{LSGI,RKMM}.
The chromodynamic interactions of heavy quarks through order $v^2/c^2$ were 
introduced in \cite{CasLep}
starting from a non relativistic treatment of QCD, defining an
effective theory called NRQCD, which was used for lattice and continuum calculations
\cite{LepThack,BBL}. This theory, together with a further effective theory derived from 
it, the potential non relativistic QCD or pNRQCD \cite{BPSV},  
is certainly one of the most diffused methods
for calculating meson spectra and decays \cite{BEHV}.
The lattice techniques, on the other hand, have 
progressively improved up to the present day by including higher relativistic orders and 
QCD radiative effects, so that increasingly accurate determinations of hyperfine
splittings have been calculated \cite{LW85,DDHH,DDHHH,LPR} . 
 
Other types of models move directly from a covariant formulation.  
A short list of up-to-date available relativistic or semi-relativistic models 
\cite{RK-CS,Bra,SeCe,HC} was discussed in our paper \cite{GSM}. 
Starting 
from our previous results \cite{GS1,GS2}, we derived  in \cite{GSM} a completely covariant equation 
for two relativistic quarks of arbitrary mass interacting  through the Cornell potential, 
with each of its two components taken  in the vector or scalar coupling appropriate for
the Dirac operators \cite{GS}.
Obviously the two spin-orbit couplings for the two quarks are completely included
in our covariant approach so no problem concerning the reduced mass  has to be posed.
A first order  correction  to the potential was added by means of the 
Breit term. We then proved that our wave equation is able to provide  a unified framework
to investigate all ranges of meson masses. 
For heavier mesons the agreement with experimental data turned out to be
really very precise up to the pair production threshold, not included in the Cornell potential.
This gave  suggestions for 
unknown spectroscopic classifications of some mesons and allowed us to obtain
a good accuracy when calculating the 
masses of light mesons, for which potential models usually fail.

All the methods so far described are  currently applied to the 
radiative meson decays, with the electromagnetic coupling  generally taken  in the 
dipole approximation (electric or
magnetic according to the considered transitions). 
Sometimes the contributions due to the strong interactions are brought 
to bear to the calculation.
In addition to the obvious comparison
of the theoretical results with experimental data, in many papers 
previsions are also made on radiative transitions lacking  direct data.
In particular this is the case of the transitions of the recently observed mesons
$h_b(1p)$ and $h_b(2p)$ decaying into $\eta_b(1s)$ and substantiating the evidence of 
$\eta_b(2s)$ \cite{Belle}.
The results of some recent papers using a semi-relativistic framework are found in
\cite{GodMoa,cinesi}, those obtained by an effective potential are in \cite{cinese}; 
QCD based approaches are used in \cite{BJV,BPV,PS1,PS2} and  
lattice calculations of three-point matrix 
elements for radiative bottomonium decays are presented in \cite{LW}.
The width of the radiative decay of $\Upsilon(2s)$
into $\eta_b(1s)$ is given in \cite{HDDHHW}.
The agreement between theoretical and experimental widths is generally not as good as 
it is for the spectra, even when taking into account the large
errors that affect the experimental data \cite{PDG}. 

The purpose of this paper is to use our two fermion
covariant potential model \cite{GSM} in order to calculate 
the purely radiative decays of  $b\bar{b}$, still assuming the Cornell potential as 
a constitutive interaction for the  mesons.
The electromagnetic coupling for the composite two fermion system is then determined in analogy to
the procedure established in \cite{BGS}, where we calculated the hyperfine spacings for 
different hydrogenic atoms and the width of corresponding transitions, without invoking
any other additional correction apart from the one given by the first order of the Breit term, 
that represents the spin-spin interaction responsible of the hyperfine splitting.
The results we found are in extremely good agreement with the experimental
data. It is therefore very tempting, if not compulsory, to have a look at the meson 
radiative decays by extending  to mesons the treatment applied in \cite{BGS} to atoms.
We thus evaluate here the branching ratios and the widths of the measured radiative decays of
$\Upsilon(3s),\,\chi_{b2}(2p),\,\chi_{b1}(2p),\,\chi_{b0}(2p),\,\Upsilon(2s)$ 
(see Tables III and IV) and we make 
previsions for some decays of $h_b(1p),\,h_b(2p),\,\chi_{b1}(2p)\Upsilon(2s),\,\Upsilon(1s)\,$
(see Table V) for which direct experimental data are not yet available. The results are 
rewarding, although we cannot expect to reach the same 
accuracy of the atomic calculations for several evident reasons.
In the first place, the Cornell potential is itself an effective
potential more suitable to the description of a stationary situation, such the calculation 
of the spectrum, as opposed to the atomic  
interaction which comes from a fundamental theory. Secondly, for atoms the fine structure
coupling constant $\alpha_{{\rm{em}}}$ is the same for the Coulomb potential, the 
Breit spin-spin interaction and the decay process. This allows us to make a proper 
calculation of the first order corrections to the wave functions due to the 
Breit term as we did in \cite{BGS}. These corrections turn out to be essential for getting very 
accurate  values of both 
the hyperfine levels of different hydrogenic atoms and the decay widths. 
Without them the levels involved in hyperfine transitions would be degenerate
and first order corrected wave functions are necessary to calculate the rate.

This is not the situation for meson radiative decays for which a 
rigorous perturbation expansion in the Breit term is
not feasible. Indeed  we have to remember that a remnant of the Breit correction 
is already present at the lowest order by means of the
parameters entering the  wave equation and its solutions.
\begin{table}[b]
	{{ $~~~$
			\begin{tabular}{lcc}
				\hline
				$~~~~~~${\texttt{State}}$\phantom{{}^{{}^{\displaystyle{i}}}}$ &$\texttt{Exp}$ 
				&$ \texttt{Num}$
				\\
				\hline
				%
				%
				%
				$({\texttt{1}}^{\texttt{1}}{\texttt{s}}_{\texttt{0}})~0^+(0^{-+})~
				\phantom{{}^{{}^{\displaystyle{i}}}}\!\! \eta_b$
				&$\phantom{XX}$\texttt{9398.0$\pm$3.2}$\phantom{XX}$  
				&\texttt{\phantom{1}9390.39}
				\\ 
				$({\texttt{1}}^{\texttt{3}}{\texttt{s}}_{\texttt{1}})~0^-(1^{--})~
				\phantom{{}^{{}^{\displaystyle{i}}}}\!\!  \varUpsilon$
				&\texttt{9460.30$\pm$.25}  
				&\texttt{\phantom{1}9466.10}
				\\ 
				%
				%
				$({\texttt{1}}^{\texttt{3}}{\texttt{p}}_{\texttt{0}})~0^+(0^{++})~
				\phantom{{}^{{}^{\displaystyle{i}}}}\!\!  \chi_{b0}$
				&\texttt{9859.44$\pm$.73}  
				&\texttt{\phantom{1}9857.41}
				\\ 
				$({\texttt{1}}^{\texttt{3}}{\texttt{p}}_{\texttt{1}})~0^+(1^{++})~
				\phantom{{}^{{}^{\displaystyle{i}}}}\!\!  \chi_{b1}$
				&\texttt{9892.78$\pm$.57}  
				&\texttt{\phantom{1}9886.70}$\spazio{0.5 } $
				\\ 
				 $({\texttt{1}}^{\texttt{1}}{\texttt{p}}_{\texttt{1}})~0^-(1^{+-})~
				 \phantom{{}^{{}^{\displaystyle{i}}}}\!\!  h_b$
				 &\texttt{9898.60$\pm$1.4}  
				 &\texttt{\phantom{1}9895.35}
				 \\
				$({\texttt{1}}^{\texttt{3}}{\texttt{p}}_{\texttt{2}})~0^+(2^{++})~
				\phantom{{}^{{}^{\displaystyle{i}}}}\!\!  \chi_{b2}$
				&\texttt{9912.21$\pm$.57}  
				&\texttt{\phantom{1}9908.14}$\spazio{0.5 } $
				\\ 
				%
				%
				$({\texttt{2}}^{\texttt{1}}{\texttt{s}}_{\texttt{0}})~0^+(0^{-+})~
				\phantom{{}^{{}^{\displaystyle{i}}}}\!\! \eta_b$
				&\texttt{\phantom{M}9974.0$\pm$4.4}${}^{\,(\ast)}$  
				&\texttt{\phantom{.}9971.14}
				\\ 
				$({\texttt{2}}^{\texttt{3}}{\texttt{s}}_{\texttt{1}})~0^-(1^{--})~
				\phantom{{}^{{}^{\displaystyle{i}}}} \!\! \varUpsilon$
				&\texttt{10023.26$\pm$.0003}  
				&\texttt{10009.04}
				\\ 
				%
				%
				%
				$({\texttt{2}}^{\texttt{3}}{\texttt{p}}_{\texttt{0}})~0^+(0^{++})~
				\phantom{{}^{{}^{\displaystyle{i}}}}\!\!  \chi_{b0}$
				&\texttt{10232.50$\pm$.0009}  
				&\texttt{10232.36}
				\\ 
				$({\texttt{2}}^{\texttt{3}}{\texttt{p}}_{\texttt{1}})~0^+(1^{++})~
				\phantom{{}^{{}^{\displaystyle{i}}}}\!\!  \chi_{b1}$
				&\texttt{10255.46$\pm$.0005}  
				&\texttt{10256.58}
				\\ 
				$({\texttt{2}}^{\texttt{1}}{\texttt{p}}_{\texttt{1}})~0^-(1^{+-})~
				\phantom{{}^{{}^{\displaystyle{i}}}}\!\!  h_b$
				&\texttt{10259.8$\pm$1.6}  
				&\texttt{\phantom{1}10263.62}
				\\
				$({\texttt{2}}^{\texttt{3}}{\texttt{p}}_{\texttt{2}})~0^+(2^{++})~
				\phantom{{}^{{}^{\displaystyle{i}}}}\!\!  \chi_{b2}$
				&\texttt{10268.65$\pm$.0007}  
				&\texttt{10274.26}
				\\
				$({\texttt{3}}^{\texttt{3}}{\texttt{s}}_{\texttt{1}})~0^-(1^{--})~
				\phantom{{}^{{}^{\displaystyle{i}}}}\!\!  \varUpsilon$
				&\texttt{10355.20$\pm$.0005}  
				&\texttt{\phantom{i}10364.52}
				$\spazio{0.5}$\\ 
				\hline
			\end{tabular}
		}}
		\label{Table_b-bbar}
		\caption{The $b\bar{b}$ levels in MeV. First column: term symbol,
			$I^G(J^{PC})$ numbers , particle name. $\sigma$=$\texttt{1.111}\,$GeV/fm,
			$\alpha$=\texttt{0.3272}, $m_b$=$\texttt{4725.5}\,$MeV. Experimental data 
			are taken from 
			\cite{PDG}.
			Our values can be compared with those obtained by different approaches, 
			reported in \cite{cinesi}.
			 ${}^{\,(\ast)}$ see \cite{Nota1}.
					}
	\end{table}
This happens because
in most quarkonium models, as in ours, the values assumed for both $\alpha_{{\rm{QCD}}}$
and for the string tension $\sigma$ are obtained from a fit of the experimental meson spectrum 
\cite{GSM},  
calculated  by taking into account the first order of the Breit correction. 
In Table I here below we report the masses of the mesons we will consider later on \cite{Nota1}.
On the other hand, in Table II we show the actually very different influence of the Breit term on 
the different states. 
 \begin{table}[t]
	{{ $~~~$
			\begin{tabular}{cccccc}
				\hline
				$~\eta_b(1s)~~~$ &$\Upsilon_b(1s)~~~$			&$\chi_{b0}(1p)~~~$		
				&$\chi_{b1}(1p)~~~$ &$h_{b}(1p)~~~$	
				&$\chi_{b2}(1p)~$
				\\
				\hline 
$~$\texttt{-92.13}$~~~$ & \texttt{-18.09}$~~~$ & \texttt{-44.3}$~~~$ & \texttt{-19.98}$~~~$ & 
\texttt{-15.95}$~~~$ & \texttt{-7.51}$~$ 
$\spazio{0.5}$\\ 
\hline
\end{tabular} 
}}
\label{Table_BreitCorr}
\caption{The Breit corrections in MeV for the lowest states.
}	\end{table} 
Due to the structure of the transition rate given in 
the following equation (\ref{w}), if we use the physical (\textit{i.e.} Breit corrected) value 
for the transition frequency it seems reasonable to take the corresponding spinors at the 
lowest perturbation order.
However for the states with $j=0$, namely $\eta_b$ and $\chi_{b0}$, the hyperfine shift is 
maximal and considerably larger than for the other states of their respective multiplets.
These states are connected by a parity transformation and 
in our model they are structurally distinguished from the
other components of the respective multiplets since they are determined by a second order
differential system instead of by a fourth order one. Moreover the inclusion of the first order
corrections in their wave functions makes a really great improvement on the results of the
decay transition rate. Still in the context of radiative meson decays, an analogous situation 
was met in \cite{Gro} for the relativistic corrections in $v^4/c^4$  ``retained in the calculation 
of those rates where those terms make a substantial difference'' (see \cite{Gro}, note 18).  
Therefore we shall
assume unperturbed wave functions for all the $j\neq 0$ states and first order corrected
wave functions for all the $j=0$ states. In the last section the numerical method of
calculating the corrections to levels and states will be recalled.

We now give a sketchy summary of what follows.
In the next section we recall the general formulae of our method: we refer for details
to our papers \cite{GSM,BGS}. 
In order to have an idea of the properties of the eigenfunctions, we
present a plot of the radial probability density of the states.
In section III we discuss some numerical aspects, describing the way
our spinors are calculated and giving some details on the numerical precision; 
we will then present the results and make some concluding remarks.
Our approach is conceptually simple;  it is completely 
covariant, so that it includes all the relativistic effects; it contains 
the Breit interaction responsible for the hyperfine splittings.
It has a general application to potential models and drastically limits the number of corrections
needed about in order to achieve agreement with experimental data. It is also rather 
manageable on the side of explicit computations in a combined environment of numerical 
methods and computer algebra.
 This is indeed the method we have used for getting  the results we have given. 
The expressions we have used and the equations we have solved can be deduced, with some work, 
from our papers \cite{GSM,BGS} by implementing the general relations to the present case which
appears simpler because of the equality of the two quark masses.
However we believe it to be useful for the reader if in Appendix  we revisit 
our method for the $b\bar b$ systems, cutting the exposition down to 
the bone and specifying only those elements necessary for the present calculations of the radiative
decay widths. This, in addition, gives us the possibility of some further observations of a more
technical character.

\sect{The transition probabilities} \label{width}
	
From \cite{GSM} we have that the two-body relativistic wave equation for $b\bar b$,
with eigenvalue $\lambda$, reads
\begin{eqnarray}
&{}&\!\!\!\Bigl[\,\Bigl(
{\gamma}^0_{(1)}{\gamma_{(1)}}_{a}-
{\gamma}^0_{(2)}{\gamma_{(2)}}_{a}\Bigr)q_a+\frac12\Bigl({\gamma}^0_{(1)}\!
+\! {\gamma}^0_{(2)}\Bigr)\Bigl(2m_b+\sigma r\Bigr)-\spazio{1.2}\cr
&{}&\phantom{XXXXX} \Bigl(\lambda+\frac b r\Bigr)+ V_B(r)
\,\Bigr]\,\Psi(\vec{r})=0.
\label{BreitHam}
\end{eqnarray}
where  $\Psi(\vec{r})$ is a 16-component spinor obtained by  reordering  the tensor 
product of the two quarks so as to collect singlets and triplets for the different
eigenvalues of the mass \cite{GS1,BGS},
${\gamma}^0_{(i)},{\gamma_{(i)}}_{a}$ are the $\gamma$ matrices acting on the space of the 
quark and anti-quark
fermion and $b=(4/3)\,\alpha_{QCD}$.
The vector and scalar parts of the Cornell potential respectively give the $(\lambda+b/r)$  and
$(2m_b+\sigma r)$ terms. Finally  the Breit potential $V_B(r)$  has the form
\begin{eqnarray}
V_B(r)=\frac b{2r}\,
{\gamma}^0_{(1)}{\gamma_{(1)}}_{a}{\gamma}^0_{(2)}{\gamma_{(2)}}_{b}
\Bigl(\delta_{ab}\!+\!\frac{r_ar_b}{r^2}\Bigr) 
\label{BreitTerm}
\end{eqnarray}
 
\begin{figure}[b]
	\centerline{\includegraphics[height=150pt]{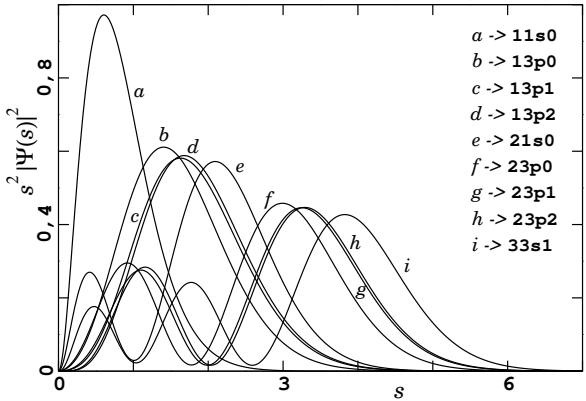}}
	\caption{\small{The normalized densities of the states involved in the $\Upsilon(3)$ decay 
			times the measure factor in the dimensionless coordinate $s$ defined in Appendix.}}
	\label{s2psi2}	
\end{figure}
	
As explained in  \cite{GSM},
when we calculate the meson spectrum the wave equation reduces to a system of four linear
differential equations (see also \cite{GS1,GS2} for further details).

We choose the reference frame with vanishing total 
momentum
$\emph{\textbf{P}}_i$ and we use the global/relative canonical coordinates $(Z,\emph{\textbf{r}})$ 
(see Appendix A of \cite{BGS}).  We factorize the wave functions of initial and final 
bottomonium states, normalized in a box of volume $V$, into
\begin{eqnarray}
\Psi_{\ell}(Z,\emph{\textbf{r}}) = V^{-1/2}\,{e^{
		-iP_{\ell}\cdot Z}}\,\psi_{\ell}(\emph{\textbf{r}})\,,\quad  \ell=i,f\,
\nonumber
\end{eqnarray}
where $\psi_i(\emph{\textbf{r}})$ and $\psi_f(\emph{\textbf{r}})$ are the 16-component spinors 
corresponding to initial and final energies, angular momenta and parities. 
The Breit term  (\ref{BreitTerm}) will always be considered a first order perturbation term. 
As previously said, when necessary we will consider wave functions represented by the 
sum of a lowest order contribution given by the exact  eigenfunction of equation (\ref{BreitHam}) 
with $V_B(r)=0$ and a first order correction generated by $V_B(r)$ itself. The general systems of 
equations for the case of mesons formed by quarks with possibly different masses is presented in 
\cite{GSM}: we report in the Appendix the systems for the $b\bar b$ case, not explicitly written
in our previous papers.
In  \cite{BGS} it was shown that the electromagnetic coupling for the fermion - anti-fermion 
bound system is introduced by means of the interaction Hamiltonian
\begin{eqnarray}
H_{\mathrm{int}} = -e_b\,\Bigl(\, \boldsymbol\alpha_{(1)}\!\cdot\!\emph{\textbf{A}}^{{(1)}}
-\boldsymbol\alpha_{(2)}\!\cdot\!\emph{\textbf{A}}^{{(2)}}\,\Bigr)
\nonumber
\end{eqnarray}
where $e_b=(1/3)\,e$ is the bottom charge, $e$ being the electron charge. Here 
$~\boldsymbol\alpha_{(i)}$ is the vector of the $\alpha$-matrices for the $i$-th
fermion space, $\emph{\textbf{A}}^{{(i)}}$ is the wave function of a photon with 4-momentum $k$ and 
polarization ${\boldsymbol\epsilon}_\sigma$ in the Coulomb gauge \cite{Landau},
$$ {\boldsymbol{A}}(k,\sigma)=
\frac{\sqrt{4\pi}}{\sqrt{2\omega\, V}}\,\,{\boldsymbol\epsilon}_\sigma\,\,e^{\displaystyle 
-i\,k_\mu x^\mu}\,,$$
calculated at the coordinate $x=x_{(i)}$, where $\omega= k_0$ is the photon frequency. 
For equal fermion masses, the standard calculation for the first perturbation order of the 
$S$-matrix element at fixed initial and final states gives
\begin{eqnarray}
&{}& S_{fi} = -ie_b\,\frac{(2\pi)^4 }{\sqrt{2\omega V}}\,\frac 
{\sqrt{4\pi}}{\sqrt{V^2}}\,\delta^4(P_f+k-P_i)\, (\boldsymbol\epsilon_\sigma^*\!\cdot\! 
\boldsymbol{M}_{fi})~~~~~
\nonumber
\label{Smatrix}
\end{eqnarray}
where
\begin{eqnarray}
\boldsymbol{M}_{fi}= \int
d^3\emph{\textbf{r}}\,\,\psi^*_{f}(\emph{\textbf{r}})\,
\Bigl(\,
\widetilde{\boldsymbol\alpha}_{(1)}
\,e^{-\frac i2{\emph{\textbf{k}}}\,\cdot\!\!\!\!{
		\emph { \textbf { r }}}}
-\widetilde{\boldsymbol\alpha}_{(2)}\,e^{\frac i2{\emph{\textbf{k}}}\,
	\cdot\!\!\!\! { \emph{ \textbf { r }}}}
\Bigr)\, \psi_{i}(\emph{\textbf{r}})\,.\spazio{0.8}
\nonumber
\end{eqnarray}
Here $\widetilde{\boldsymbol\alpha}_{(j)}$ are the transformed matrices in the basis of the
spinors with reordered components above mentioned.
The $\delta^4$-function gives the conservation of the global 4-momentum 
$$P^\mu_i = P^\mu_f + k^\mu$$ and contains the recoil
of the meson due to the radiation emission.
Using the explicit form  of the spinors obtained by the diagonalization of total angular momentum 
and parity (see Appendix), the integrals over the angular variables can
be analytically calculated, without any approximation on the exponential, in terms of elementary 
functions that correspond to the lowest order Bessel functions appearing in the usual series 
expansions. The radial integrals are then calculated numerically for all the allowed transitions 
as described in more detail in the next section.

Summing over the polarizations and the possible final states and averaging over the initial states, 
the differential transition rate therefore reads
\begin{eqnarray}
dw =
\frac{e_b^2}{2\pi\omega}\,\delta^4(P_f+k-P_i)\sum\,
\frac{|\boldsymbol\epsilon_\sigma^*\!\cdot\!\!\emph{\textbf{M}}_{fi}|^2}{2j_i+1}
\,\,{d^3\!\!\emph { \textbf{k}}}\,\,{d^3\emph{\textbf{P}}_f}
\spazio{1.2}
\nonumber
\end{eqnarray}
As
$\int{d^3\emph{\textbf{P}}}/{2P^0} = \int d^4{P}\,\theta(P^0)\,\delta(P^2-\lambda^2)\,,$
integrating over the final global momentum we get
\begin{eqnarray}
\frac{dw}{d\omega\,d\Omega_n}\! = \!\frac{e_b^2\,\omega
}{2\pi\lambda_i}\,
(\lambda_i-\omega)\,\delta\Bigl(\omega-\frac{\lambda_i^2-\lambda_f^2}{2\lambda_i}\Bigr)
\sum\,
\frac{|\boldsymbol\epsilon_\sigma^*\!\cdot\!\!\emph{\textbf{M}}_{fi}|^2}{2j_i+1} 
\nonumber
\end{eqnarray}
where $d\Omega_n$ is the unit solid angle in the direction 
$\emph{\textbf{n}}=\emph{\textbf{k}}/|\emph{\textbf{k}}|$.
Reinserting the $\hbar$ and $c$ factors, the final integration over the solid angle gives the total 
transition rate
\begin{eqnarray}
w = \frac 43\,\frac{e_b^2}{\hbar c}\,\,\omega_{fi}\,
\,\Lambda_{fi}^2\,\sum\,
\frac{|\boldsymbol\epsilon_\sigma^*\!\cdot\!\!\emph{\textbf{M}}_{fi}|^2}{2j_i+1}  
\label{w}
\end{eqnarray}
while
\begin{eqnarray}
\omega_{fi}= \frac c\hbar\,\,\frac{\lambda_i+\lambda_f}{2\lambda_i}\,(\lambda_i-\lambda_f)
\nonumber
\end{eqnarray}
is the frequency of the emitted photon that completely includes the recoil. Finally 
\begin{eqnarray}
\Lambda_{fi}^2 = \frac{\lambda^2_i+\lambda^2_f}{2\lambda^2_i}
\nonumber
\end{eqnarray}
is the relativistic correction factor coming from kinematics. This is not 
very far from unity but for the transitions between states with large mass difference.



\section{NUMERICAL RESULTS AND CONCLUSIONS} \label{conclusions}


As extensively described in \cite{GSM,BGS}, the levels and 
the eigenstates are  obtained by a numerical solution of the singular boundary value 
problem (BVP) posed by the Hamiltonian (\ref{BreitHam}) acting on 16-component spinors 
involving eight radial function coefficients 
$\{a_i(r),b_i(r),c_i(r),d_i(r)\}_{i=0,1}$. The symmetries of the problem imply four algebraic 
relations, so that the previous radial functions can be expressed in terms of only four
unknown functions $\{u_i(r)\}_{i=1,4}$ and the the BVP reduces to a $4\times4$ linear differential 
system for each parity. 
As stated above, for the $j=0$ states, the $4\times4$ system actually reduces to a $2\times2$ one.
In the Appendix we specify the matrix elements of the reduced differential 
systems and we give the explicit  expression of the $a_i,b_i,c_i,d_i $ in terms of the $u_k$.
We also add some comments on the $P$- and $PC$-transformations. 
\begin{table}[t]
	{{ $~~~$
			\begin{tabular}{lrr}
				\hline	
				\texttt{\phantom{XXXXXXXX}Branching Ratios}
				&  \texttt{Theor}  
				&  \texttt{Exp\phantom{X.}}	$\spazio{0.5}$\\
				\hline  
				$\Upsilon(3s)\rightarrow\gamma\chi_{b1}(2p)/\Upsilon(3s)
				\rightarrow\gamma\chi_{b2}(2p)~~ $
				&  \texttt{.812}$~$  
				&  \texttt{.96$\pm$.21}  	
				\\
				$\Upsilon(3s)\rightarrow\gamma\chi_{b0}(2p)/\Upsilon(3s)
				\rightarrow\gamma\chi_{b2}(2p)~~ $
				&  \texttt{.433}$~$  
				&  \texttt{.45$\pm$.10}  	
				\\
				$\Upsilon(3s)\rightarrow\gamma\eta_{b}(2s)/\Upsilon(3s)
				\rightarrow\gamma\chi_{b2}(2p)~~ $
				&  \texttt{.002}$~$  
				&  \texttt{<.005\phantom{X}}  	
				\\
				$\Upsilon(3s)\rightarrow\gamma\chi_{b2}(1p)/\Upsilon(3s)
				\rightarrow\gamma\chi_{b2}(2p)~~ $
				&  \texttt{.042}$~$  
				&  \texttt{.075$\pm$.019}  	
				\\
				$\Upsilon(3s)\rightarrow\gamma\chi_{b1}(1p)/\Upsilon(3s)
				\rightarrow\gamma\chi_{b2}(2p)~~ $
				&  \texttt{.010}$~$  
				&  \texttt{.007$\pm$.005}  	
				\\
				$\Upsilon(3s)\rightarrow\gamma\chi_{b0}(1p)/\Upsilon(3s)
				\rightarrow\gamma\chi_{b2}(2p)~~ $
				&  \texttt{.010}$~$  
				&  \texttt{.021$\pm$.006}  	
				\\	
				$\Upsilon(3s)\rightarrow\gamma\eta_{b}(1s)/\Upsilon(3s)
				\rightarrow\gamma\chi_{b2}(2p)~~ $
				&  \texttt{.003}$~$  
				&  \texttt{.004$\pm$.001}$\!\!\!$  				
				$\spazio{0.75}$\\	
				\hline 	
				$\Upsilon(2s)\rightarrow\gamma\chi_{b1}(1p)/\Upsilon(2s)
				\rightarrow\gamma\chi_{b2}(1p)~~ $
				&  \texttt{.812}$~$    
				&  \texttt{.96$\pm$.10}  	
				\\
				$\Upsilon(2s)\rightarrow\gamma\chi_{b0}(1p)/\Upsilon(2s)
				\rightarrow\gamma\chi_{b2}(1p)~~ $
				&  \texttt{.410}$~$  
				&  \texttt{.53$\pm$.08}  	
				\\
				$\Upsilon(2s)\rightarrow\gamma\eta_{b}(1s)/\Upsilon(2s)
				\rightarrow\gamma\chi_{b2}(1p)~~ $
				&  \texttt{.006}$~$  
				&  \texttt{.006$\pm$.002}$\!\!\!$  	
				$\spazio{0.75}$\\
				\hline
				$\chi_{b2}(2p)\rightarrow\gamma\Upsilon(1s)/\chi_{b2}(2p)\rightarrow\gamma\Upsilon(2s)~~
				$
				&  \texttt{.55}$~$  
				&  \texttt{.66$\pm$.23}$\!\!\!$ 	
				$\spazio{0.75}$\\
				\hline					
				$\chi_{b1}(2p)\rightarrow\gamma\Upsilon(1s)/\chi_{b1}(2p)\rightarrow\gamma\Upsilon(2s)~~
				$
				&  \texttt{.46}$~$  
				&  \texttt{.46$\pm$.08}$\!\!\!$ 	
				$\spazio{0.75}$\\
				\hline					
				$\chi_{b0}(2p)\rightarrow\gamma\Upsilon(1s)/\chi_{b0}(2p)\rightarrow\gamma\Upsilon(2s)~~
				$
				&  \texttt{.13}$~$  
				&  \texttt{(.20$\pm$.20)}${}^{(\ast)}$ $\!\!\!\!\!\!$  	
				$\spazio{0.75}$\\
				\hline					
				
			\end{tabular}
		}}
		\label{Table_BR}
		\caption{Comparison of our calculated branching ratios of the radiative decays with 
			experimental data. The experimental errors have been linearly combined. 
			${}^{(\ast)}$This 
			case 
			is
			not very meaningful due to the large error by which it is affected.
		}
	\end{table}
	%
%
\begin{table}[h]
	{{ $~~~$
			\begin{tabular}{lrr}
				\hline
				$~~~~$\texttt{Decay}  
				&  \texttt{\phantom{xx}Theor}  
				&  \texttt{Exp\phantom{XX}}	$\spazio{0.5}$\\
				\hline 
				%
				%
				%
				$\Upsilon(3s)
				\rightarrow\gamma\chi_{b2}(2p)~~ $
				%
				
				%
				&$~~~$\texttt{3.51 }$~$  
				&$~~$\texttt{2.70$\pm$0.57 }
				\\ 
				$\Upsilon(3s)
				\rightarrow\gamma\chi_{b1}(2p)~~ $%
				
				%
				&$~~~$\texttt{2.85 }$~$
				&$~~$\texttt{2.58$\pm$0.48  }$\!\!\!$
				$\spazio{0.5}$\\
				%
				%
				$\Upsilon(3s)
				\rightarrow\gamma\chi_{b0}(2p)~~ $%
				
				%
				&$~~~$\texttt{1.52 }$~$  
				&$~~$\texttt{1.21$\pm$0.23  } 
				\\ 
				$\Upsilon(3s)
				\rightarrow\gamma\eta_{b}(2s)~~ $%
				
				%
				&$~~~$\texttt{0.006 }$~$		 
				&$~~$\texttt{<$\,$0.013  }	 
				\\ 
				$\Upsilon(3s)
				\rightarrow\gamma\chi_{b2}(1p)~~ $	%
				
				%
				&$~~~$\texttt{0.149 }$~$  
				&$~~$\texttt{0.204$\pm$0.045  }
				\\ 
				%
				%
				$\Upsilon(3s)
				\rightarrow\gamma\chi_{b1}(1p)~~ $	%
				
				%
				&$~~~$\texttt{0.036 }$~$  
				&$~~$\texttt{0.019$\pm$0.012  } 
				\\ 
				$\Upsilon(3s)
				\rightarrow\gamma\chi_{b0}(1p)~~ $	%
				
				%
				&$~~~$\texttt{0.032 }$~$  
				&$~~$\texttt{$\phantom {.}$0.056$\pm$0.013  }$\!\!\!$   
				$\spazio{0.5}$\\
				%
				%
				%
				$\Upsilon(3s)
				\rightarrow\gamma\eta_{b}(1s)~~ $%
				
				%
				&$~~~$\texttt{0.009 }$~$  
				&$~~$\texttt{0.011$\pm$0.003  }$\!\!\!$  
				$\spazio{0.5}$\\ 
				\hline
				$\Upsilon(2s)
				\rightarrow\gamma\chi_{b2}(1p)~~ $	%
				
				%
				&$~~~$\texttt{2.13 }$~$  
				&$~~$\texttt{2.30$\pm$0.20  }$\!\!$
				\\ 
				%
				%
				$\Upsilon(2s)
				\rightarrow\gamma\chi_{b1}(1p)~~ $	%
				
				%
				&$~~~$\texttt{1.73 }$~$  
				&$~~$\texttt{2.22$\pm$0.21  }$\!\!\!$ 
				\\ 
				$\Upsilon(2s)
				\rightarrow\gamma\chi_{b0}(1p)~~ $	%
				
				%
				&$~~~$\texttt{0.87 }$~$  
				&$~~$\texttt{$\phantom {.}$1.22$\pm$0.15  } $\!\!\!\!\!\!\!$  
				$\spazio{0.5}$\\
				%
				%
				%
				$\Upsilon(2s)
				\rightarrow\gamma\eta_{b}(1s)~~ $%
				
				%
				&$~~~$\texttt{0.013 }$~$  
				&$~~$\texttt{0.013$\pm$0.04  }$\!\!\!\!$  
				$\spazio{0.5}$\\ 
				\hline
				$\chi_{b2}(2p)
				\rightarrow\gamma\Upsilon(2s)~~ $%
				%
				&$~~~$\texttt{18.77 }$~$  
				&$~~$\texttt{15.10$\pm$5.60 }$\!\!$  
				\\  
				$\chi_{b2}(2p)
				\rightarrow\gamma\Upsilon(1s)~~ $%
				%
				&$~~~$\texttt{10.27 }$~$  
				&$~~$\texttt{9.80$\pm$2.30 }$\!\!\!\!$
				$\spazio{0.5}$\\ 
				\hline
				$\chi_{b1}(2p)
				\rightarrow\gamma\Upsilon(2s)~~ $%
				%
				&$~~~$\texttt{16.80 }$~$  
				&$~~$\texttt{14.40$\pm$5.00 }$\!\!\!$
				\\  
				$\chi_{b1}(2p)
				\rightarrow\gamma\Upsilon(1s)~~ $%
				%
				&$~~~$\texttt{7.68 }$~$  
				&$~~$\texttt{8.96$\pm$2.24 }$\!\!\!\!$
				$\spazio{0.5}$\\ 
				\hline	
				$\chi_{b0}(2p)
				\rightarrow\gamma\Upsilon(2s)~~ $%
				%
				&$~~~$\texttt{11.77 }$~$  
				&$~~$\texttt{- }$\phantom {XX}$
				\\  
				$\chi_{b0}(2p)
				\rightarrow\gamma\Upsilon(1s)~~ $%
				%
				&$~~~$\texttt{1.49 }$~$  
				&$~~$\texttt{- }$\phantom {XX}$	\\
				\hline
				$\chi_{b2}(1p)
				\rightarrow\gamma\Upsilon(1s)~~ $%
				%
				&$~~~$\texttt{33.73 }$~$  
				&$~~$\texttt{- }$\phantom {XX}$	\\
				\hline	
				$\chi_{b1}(1p)
				\rightarrow\gamma\Upsilon(1s)~~ $%
				%
				&$~~~$\texttt{29.48 }$~$  
				&$~~$\texttt{- }$\phantom {XX}$	\\
				\hline	
				$\chi_{b0}(1p)
				\rightarrow\gamma\Upsilon(1s)~~ $%
				%
				&$~~~$\texttt{19.65 }$~$  
				&$~~$\texttt{- }$\phantom {XX}$	\\
				\hline		
			\end{tabular}
		}}
		\label{Table_beta}
		\caption{Values in keV of the theoretical and experimental widths of radiative decays 
			of $\Upsilon(3s),\,\chi_{b2}(2p)\,,\chi_{b1}(2p)\,,\chi_{b0}(2p)\,,\Upsilon(2s)$.
		}
	\end{table}
 Moreover, the corrections 
 due to the Breit term (\ref{BreitTerm}) are calculated by solving the BVP for $\varepsilon V_B(r)$ 
 and taking the first order expansion in $\varepsilon$ both for levels and spinors
 normalized to unity.
 Due to the value 
 of the bottom mass, the parameters entering the BVP are such that the solutions can
 be expressed directly in terms of Pad\'e approximants, so that we get a complete control of 
 the numerical error. The Pad\'e we have calculated are of order [260,260] and since the 
 arithmetical precision has also been taken with a sufficiently large number of digits, our 
 numerical errors 
can safely be assumed to be vanishing. Just to have a consistency test of our eigenfunctions with
the covariant framework in which they are determined, we have calculated some estimates of the
average meson radii and quark velocity. For the radii we have seen a monotonic increase with mass, 
from $0.156\,\, $fm of the $\eta_b(1s)$ to $0.609\,\, $fm of  $\Upsilon(3p)$. The estimate of the 
quark 
velocity has been approximated by $\beta \simeq (\langle 
{\boldsymbol{q}}^2\rangle/(m^2c^2+\langle {\boldsymbol{q}}^2\rangle))^{1/2}$, where $q=(p_1-p_2)/2$ 
is the conjugate relative momentum of the two fermions 
\cite{GSM,BGS} and  $\langle {\boldsymbol{q}}^2\rangle$  has been taken as the he average of 
$\,-\hbar^2\,{\boldsymbol{\nabla}}^2_r\,$ over the corresponding state. Averaging then over all the 
states we have found a $\beta\simeq 0.29$ in good agreement with the estimate of \cite{PS2}. 
Such a value, in our opinion,  strengthens the idea that a completely relativistic
treatment is appropriate. Finally we have calculated the average values of the orbital 
angular momentum from the radial part of the Laplace operator and we have found values no larger
than 0.115 for $s$-states and values between 0.976 and 1.045 for $p$-states.

In Table III we present our results for `relative' branching ratios of  bottomonium radiative 
decays, namely for branching ratios not using the total width of the decaying particle and we 
make a comparison with experimental data whose errors have been linearly combined. It appears that
the agreement between theoretical and experimental data is very good for most of the decays and that
the worst results are different for a factor not greater than 1.5.

%

In the second column of Table IV we report the values in keV we have calculated for the $b\bar b$ 
mesons radiative decays. We get from \cite{PDG} the total widths 
$\Gamma_{\Upsilon(3s)}=20.32\pm1.85\,\,$keV, $\Gamma_{\Upsilon(2s)}=31.98\pm2.63\,\,$keV, while the
total widths of $\Gamma_{\chi_{b2}(2p)}=138.\pm19.\,\,$keV and 
$\Gamma_{\chi_{b1}(2p)}=96.\pm16.\,\,$keV are taken from \cite{Caw}. Again we have assumed a linear
composition of the errors of the experimental data.
The agreement is again very good even for the decays involving the $\eta_b$ and the
$\chi_{b0}$ states. Tables of the results for the decay withs
obtained by several different approaches can be found in \cite{cinesi}.

In Table V we give the results of our previsions in comparison with those of References
\cite{cinesi} and \cite{PS2}.


\section*{Appendix} \label{appendix}
%
The general notations, the derivation of the wave equation and the proof of
many of its properties (variables, covariance and so on) have been given in \cite{GS1,GS2,GSM}.
The eigenfunctions of the Hamiltonian (\ref{BreitHam}), 
obtained by coupling two Dirac equations, are 16-dimensional spinors.

If  $m_b$ is the bottom mass, we introduce the 
dimensionless variables $\Omega,\,w,\,s$ by
\begin{eqnarray}
\!\!\!\sigma=m_b^2\,\Omega^{\frac32},~~ \lambda=m_b\,(2+\Omega w),~~ 
r=m_b^{-1}\,\Omega^{-\frac12}\,s,
\label{Dimensionless_Variables}
\nonumber 
\end{eqnarray}
$\blacktriangleright$  The parity operator $P$ is given by  the inner parity -- which is
the reordered $\gamma_0\otimes\gamma_0$ -- combined with the change $\boldsymbol{r}
\rightarrow -\boldsymbol{r}$ times a global arbitrary phase. In our previous papers dealing
with atomic states  we had called ``even'' or ``odd'' the states with eigenvalues of $P$
equal to ${(-)}^j$ or ${(-)}^{j +1}$ respectively, choosing the arbitrary phase equal to unity.
With such a choice the ground state of an atomic system has the $P$ eigenvalue equal to unity. 
We maintain this terminology but we observe that we have to choose the global phase
equal to -1 in order to agree with the usual meson classification scheme. 

\begin{table}[t]
	{{ $~~~$
			\begin{tabular}{lrrr}
				\hline
				$~~~~$\texttt{Decay} 
				&  \texttt{Ours} $~~~$ 
				& \texttt{Ref[28]} $~$ 
				& \texttt{Ref[23]} $~$  	$\spazio{0.5}$\\
				\hline 
				%
				%
				%
				
				%
				$ h_{b}(2p)
				\rightarrow\gamma\eta_{b}(2s)~~ $ $~~$%
				%
				&\texttt{20681} $~~$ 
				&\texttt{17600} $~~$ 
				&\texttt{16600} $~~$ 
				\\
				%
				$ h_{b}(2p)
				\rightarrow\gamma\eta_{b}(1s)~~ $ $~~$%
				%
				&\texttt{16884} $~~$ 
				&\texttt{14900} $~~$ 
				&\texttt{17500} $~~$ 
				\\
				\hline
				$\Upsilon(2s)
				\rightarrow\gamma\eta_{b}(2s)~~ $ $~~$%
				%
				&\texttt{0.369} $~~$
				& \texttt{0.58} $~~$ 
				&\texttt{0.59} $~~$
				$\spazio{0.5}$\\ 
				\hline 
				$\eta_b(2s)
				\rightarrow\gamma\Upsilon(1s)~~ $ $~~$%
				%
				&\texttt{65.41} $~~$
				& \texttt{45} $~~$ 
				&\texttt{64} $~~$
				$\spazio{0.5}$\\ 	
				\hline	
				$\chi_{b2}(1p)
				\rightarrow\gamma h_{b}(1p)~~ $ $~~$
				%
				&\texttt{0.015} $~~$
				&\texttt{0.089} $~~$		 
				&\texttt{  } $~~$	 
				\\ 
				$\chi_{b2}(1p)
				\rightarrow\gamma \Upsilon(1s)$ $~~$
				%
				&
				\texttt{33731} $~~$	
				&\texttt{39150} $~~$	 
				&\texttt{31800} $~~$	 
				\\
				\hline
				$ h_{b}(1p)
				\rightarrow\gamma\chi_{b1}(1p)~~ $ $~~$%
				%
				&\texttt{0.050} $~~$ 
				& \texttt{0.012} $~~$
				&\texttt{0.0094} $~~$
				\\ 
				%
				%
				$ h_{b}(1p)
				\rightarrow\gamma\chi_{b0}(1p)~~ $ $~~$	%
				
				%
				&\texttt{0.124} $~~$  
				&\texttt{0.86} $~~$
				&\texttt{0.90} $~~$ 
				\\ 
				$ h_{b}(1p)
				\rightarrow\gamma\eta_b(1s)~~ $ $~~$	%
				%
				&\texttt{39318} $~~$ 
				& \texttt{43660} $~~$
				&\texttt{35800} $~~$ 
				\\ 
				\hline
				$\Upsilon(1s)
				\rightarrow\gamma\eta_{b}(1s)~~ $ $~~$%
				%
				&\texttt{3.101${}^{\boldsymbol{\ast}}$} $~~$  
				&\texttt{9.34} $~~$
				&$~~$\texttt{10} $~~$  
				$\spazio{0.5}$\\
				\hline
				
			\end{tabular}
		}}
		\label{Table_beta}
		\caption{Comparison of the previsions for the theoretical widths of some radiative 
			decays 
			of $\chi_{b2},\,h_b,\,\chi_{b1},\,\chi_{b0}\,\,{\mathrm{ and}} \,\,\Upsilon$. Units 
			are eV. 	${}^\ast$ This value is in agreement with the value (3.6$\pm$2.9)$\,$eV 
			of \cite{BPV}.
		}
	\end{table}
	
After diagonalization of angular momentum and parity the components of the 
general reordered spinor we have called ``even'' are \cite{GS1,BGS}
\begin{eqnarray} 
%
%
%
&{}& \psi_{+\,0}^{(M)}= Y^{j}_{m}(\theta,\,\phi)\,{a_{0}}(s)
~~~~~~~~~\, 
\spazio {1.2}  \cr
%
%
&{}& \psi_{+\,1_+}^{(M)}= - 
{\displaystyle \frac {\sqrt{j - m + 1}\sqrt{j + m}}{\sqrt{2\,j}
		\sqrt{j+1}}}\,
Y^{j}_{m-1}(\theta,\,\phi)\,{b_{0}}(s)   
\spazio {1.2}  \cr
%
%
&{}&\psi_{+\,1_0}^{(M)}= {\displaystyle \frac {m}{\sqrt{j}\,\sqrt{1 + j}}}
\,Y^{j}_{m}(\theta,\,\phi)\,
\,{b_{0}}(s) 
\spazio {1.2}  \cr
%
%
&{}&\psi_{+\,1_-}^{(M)}= 
{\displaystyle \frac {
		\sqrt{j - m}\,\sqrt{j + m + 1}}{\sqrt{2\,j}
		\sqrt{j+1}}}\,
Y^{j}_{m+1}(\theta,\,\phi)\,{b_{0}}(s) 
\spazio {1.2}  \nonumber\cr
%
%
%
%
&{}& \psi_{+\,0}^{(-M)}=Y^{j}_{m}(\theta,\,\phi)\,{a_{1}}(s)
\spazio {1.2}  \cr
%
%
&{}& \psi_{+\,1_+}^{(-M)}=- {\displaystyle 
	\frac {\sqrt{j - m + 1}\sqrt{j + m}}{\sqrt{2\,j}\sqrt{
			j+1}}}\,
Y^{j}_{m-1}(\theta,\phi)\,{b_{1}}(s)    
\spazio {1.2}  \cr
%
%
&{}&\psi_{+\,1_0}^{(-M)}={\displaystyle \frac {m}{\sqrt{j}\,\sqrt{1 + j}}}
\,Y^{j}_{m}(\theta,\,\phi)
\,{b_{1}}(s)
\spazio {1.2}  \cr
%
%
&{}&\psi_{+\,1_-}^{(-M)}= {\displaystyle \frac {\sqrt{j - m}\sqrt{j + m + 1}}{\sqrt{2\,j}
		\sqrt{j+1}}}\,
Y^{j}_{m+1}(\theta,\phi)\,{b_{1}}(s)   
\spazio {1.2}  \nonumber\cr
%
%
%
%
%
%
&{}& \psi_{+\,0}^{(-\mu)}=0
\spazio {1.2}  \cr
%
%
&{}& \psi_{+\,1_+}^{(-\mu)}={\displaystyle 
	\frac {\sqrt{j + m
			- 1}\,\sqrt{j + m}}{\sqrt{2\,j}\,\sqrt{2\,j - 1}}}\,
Y^{j-1}_{m-1}(\theta,\,\phi)\,
{c_0}(s) + \spazio {1.2}  \cr
&{}& \phantom{\psi_{+\,1_+}^{(-\mu)}=X}{\displaystyle \frac {\sqrt{j - m + 1}\,
		\sqrt{j - m + 2
		}}{
		\sqrt{ 2\,j+2}\,\sqrt{ 2\,j+3}}}
\,Y^{j+1}_{m-1}(\theta,\,\phi)\,{d_{0}}(s) 
\spazio {1.2}  \cr
%
%
&{}&\psi_{+\,1_0}^{(-\mu)}={\displaystyle \frac {\sqrt{j - m}\,\sqrt{j + m}}{\sqrt{j}\,
		\sqrt{2\,j
			- 1}}}\,
Y^{j-1}_{m}(\theta,\,\phi)\,
{c_0}(s) -\spazio {1.2}  \cr
&{}& \phantom{\psi_{+\,1_0}^{(-\mu)}=X}{\displaystyle \frac {\sqrt{j - m + 1}\,
		\sqrt{j + m + 1
		}}{\sqrt{
		1 + j}\,\sqrt{ 2\,j+3}}}\,Y^{j+1}_{m}(\theta,\,\phi)\,{d_{0}}(s)  
\spazio {1.2}  \cr
%
%
&{}&\psi_{+\,1_-}^{(-\mu)}={\displaystyle \frac {\sqrt{j - m
			- 1}\,\sqrt{j - m}}{\sqrt{2\,j}\,\sqrt{2\,j - 1}}}\,
Y^{j-1}_{m+1}(\theta,\,\phi)\,
{c_0}(s) +\spazio {1.2}  \cr
&{}& \phantom{\psi_{+\,1_-}^{(-\mu)}=X}{\displaystyle \frac {\sqrt{j + m + 1}\,
		\sqrt{j + m + 2
		}}{
		\sqrt{ 2\,j+2}\,\sqrt{ 2\,j+3}}}
\,Y^{j+1}_{m+1}(\theta,\,\phi)\,{d_{0}}(s)   
\spazio {1.2}  \nonumber\cr
%
%
%
%
&{}& \psi_{+\,0}^{(\mu)}=0  
\spazio {1.2}  \cr
%
%
&{}&\psi_{+\,1_+}^{(\mu)}={\displaystyle 
	\frac {\sqrt{j + m
			- 1}\,\sqrt{j + m}}{\sqrt{2\,j}\,\sqrt{2\,j - 1}}}\,
Y^{j-1}_{m-1}(\theta,\,\phi)\,
{c_{1}}(s) + \spazio {1.2}  \cr
&{}& \phantom{\psi_{+\,1_+}^{(-\mu)}=X}{\displaystyle \frac {\sqrt{j - m + 1}\,
		\sqrt{j- m + 2
		}}{
		\sqrt{ 2\,j+2}\,\sqrt{ 2\,j+3}}}
\,Y^{j+1}_{m-1}(\theta,\,\phi)\,{d_{1}}(s)   
\spazio {1.2}  \cr
%
%
&{}&\psi_{+\,1_0}^{(\mu)}={\displaystyle \frac {\sqrt{j - m}\,\sqrt{j + m}}{\sqrt{j}
		\,\sqrt{2\,j - 1
		}}}\,
		Y^{j-1}_{m}(\theta,\,\phi)\,
		{c_{1}}(s) - \spazio {1.2}  \cr
	&{}& \phantom{\psi_{+\,1_0}^{(-\mu)}=X}{\displaystyle \frac {\sqrt{j - m + 1}\,
				\sqrt{j + m + 1
				}}{\sqrt{
				j + 1}\,\sqrt{ 2\,j+3}}}
	\,Y^{j+1}_{m}(\theta,\,\phi)\,{d_{1}}(s)   
	\spazio {1.2}  \cr
	%
	%
	&{}&\psi_{+\,1_-}^{(\mu)}={\displaystyle 
		\frac {\sqrt{j - m
				- 1}\,\sqrt{j - m}}{\sqrt{2\,j}\,\sqrt{2\,j - 1}}}\,
	Y^{j-1}_{m+1}(\theta,\,\phi)\,
	{c_{1}}(s) +  \spazio {1.2}  \cr
	&{}& \phantom{\psi_{+\,1_-}^{(-\mu)}=X}{\displaystyle \frac {\sqrt{j + m + 1}\,
			\sqrt{j + m + 2
			}}{
			\sqrt{ 2\,j+2}\,\sqrt{ 2\,j+3}}}
	\,Y^{j+1}_{m+1}(\theta,\,\phi)\,{d_{1}}(s)
	\nonumber
	%
	%
	%
	%
	\label{psipari16}
	\end{eqnarray}
The state with opposite parity is obtained by applying to the previous spinor 
the block matrix with the $8\times 8$ zero 
matrices on the diagonal and the $8\times 8$ identity matrices on the anti-diagonal.

$\blacktriangleright$ The coefficient functions $\{a_i(s),b_i(s),c_i(s),d_i(s)\}_{i=0,1}$ for the 
eigenstates
are obtained by solving a boundary value problem which, due to the symmetries of the Hamiltonian is 
actually equivalent to the solution of a reduced $4\times 4$ system for each different parity.
The system is \cite{GSM}
\begin{eqnarray}
\!\left(\!\! {\begin{array}{c}
	u'_1(s)\\
	u'_2(s)\\
	u'_3(s)\\
	u'_4(s)
	\end{array}}\!\!\right)\!\!+\!\! 
\left( \!\!{\begin{array}{cccc} 
	0 &\!\!\!\phantom{-}A_0(s) &\!\!\!\,\, -B_0(s) &\!\!\! \phantom{-}0\\
	A_\varepsilon(s) &\!\!\! \phantom{-}{ {1}/{s}}  &\!\!\! \phantom{-}0 &\!\!\!   B_\varepsilon(s) 
	\\ 
	{C}_\varepsilon(s) &\!\!\!\phantom{-}0 &\!\!\! \phantom{-} { {2}/{s}}  &\!\!\! 
	\phantom{-}A_\varepsilon(s)\\ 
	0 &\!\!\!  {D}_\varepsilon(s) &\!\!\!\phantom{-}A_0(s) &\!\!\! \phantom{-}{ {1}/{s}}
	\end{array}}
\!\! \right)\!\!
\left(\!\! {\begin{array}{c}
	u_1(s)\\
	u_2(s)\\
	u_3(s)\\
	u_4(s)
	\end{array}}\!\!\right)\!= 0.
\label{System}
\nonumber
\end{eqnarray}
with $A_0\!=\!A_\varepsilon|_{\varepsilon=0}$,
$B_0\!=\!B_\varepsilon|_{\varepsilon=0}\,$ and where the four unknown functions $\{u_i(s)\}_{i=1,4}$
determine the above eight $\{a_i(s),b_i(s),c_i(s),d_i(s)\}_{i=0,1}$.
The ``even'' parity coefficients  for $b\bar b$ are
\begin{eqnarray}
&{}&A_\varepsilon(s)= 0
\spazio{1.2}\cr
&{}& B_\varepsilon(s)= \frac1{2s}\,(sh(s)+2\varepsilon b)
\,,\spazio{1.2}\cr 
&{}&C_\varepsilon(s)=\frac1{2s}\,(sh(s)+2\varepsilon b)-
\frac{2J^2}{s(sh(s)-2\varepsilon b)}-\frac{2s\,k^2(s)}{s\,h(s)-4\varepsilon b}
\,\spazio{1.2}\cr
&{}&D_\varepsilon(s)= -\frac1{2s}\,(sh(s)+2\varepsilon b)+\frac{2J^2}{s^2h(s)}
+\frac{2s\,k^2(s)}{s\,h(s)-2\varepsilon b}
\label{CFeven}
\nonumber
\end{eqnarray} 
%
%
$\blacktriangleright$ The ``odd'' parity coefficients for $b\bar b$ read \cite{Nota2} 
\begin{eqnarray}
&{}&A_\varepsilon(s)= -\frac{2\,\sqrt{J^2}\,k(s)}{s\,h(s)-2\varepsilon b},
\spazio{1.0}\cr
&{}&B_\varepsilon(s)=\frac1{2s}\,(sh(s)+2\varepsilon b)-\frac{2s\,k^2(s)}{sh(s)-2\varepsilon b}
	\,\spazio{1.0}\cr
&{}&C_\varepsilon(s)=\frac1{2s}\,(sh(s)+4\varepsilon b)-\frac{2J^2}{s(sh(s)-2\varepsilon b)}
\spazio{1.0}\cr
&{}&D_\varepsilon(s)=-\frac{h(s)}2+\frac{2J^2}{s^2h(s)}-\frac{\varepsilon b}{s}
\nonumber
\label{CFodd}
\end{eqnarray}
In (\ref{CFeven},\ref{CFodd}) $J^2=j(j+1)$ and
\begin{eqnarray}
h(s)=(2+\Omega w)/\sqrt{\Omega}+ b/s,\,\,  k(s)=(2+\Omega s)/(2\sqrt{\Omega})\,.
\nonumber
\end{eqnarray} 
$\blacktriangleright$ For integer $n$ we let
\begin{eqnarray}
\Delta_n(s) = (2+\Omega\,w)\,s+\sqrt{\Omega}\,b\,(1-n\varepsilon)
\nonumber
\end{eqnarray}
The relations among the four $u_i$ and the eight $a_i,b_i,c_i,d_i$ 
variables for the ``even'' states are then
\begin{eqnarray}
&{}&\displaystyle 
	a_0(s)=\frac12\, \left( 1+{\frac { \left( \Omega\,s+2 \right) s}{{\Delta_4(s)}}} 
	\right)\mbox{}u_{{1}} \left( s \right) \spazio{1.2}\cr
&{}&\displaystyle 
	a_1(s)=\frac12\, \left( 1-{\frac { \left( \Omega\,s+2 \right) s}{{\Delta_4(s)} }} 
	\right) \mbox{}u_{{1}} \left( s \right) \spazio{1.2}\cr
&{}&\displaystyle 
	b_0(s)=\frac12\, \left( 1+{\frac { \left( \Omega\,s+2 \right) s}{{\Delta_2(s)}}} 
	\right)\mbox{}u_{{2}} \left( s \right)\spazio{1.2}\cr
&{}&\displaystyle 
    b_1(s) =\frac12\, \left( -1+{\frac { \left( \Omega\,s+2 \right) s}{{\Delta_2(s)} 
	}} \right) \mbox{}u_{{2}} \left( s \right)\spazio{1.2}\cr
&{}& c_0(s) =-{\frac { \sqrt{\Omega} \left( j+1 \right)  \sqrt{j}\,u_{{1}} 
 		\left( s \right) }{ \sqrt{2\,j+1}\,{\Delta_2(s)} \mbox{}}}+{\frac { \sqrt{j+1} 
 		\sqrt{\Omega}j\,u_{{2}} \left( s \right) }{ \sqrt{2\,j+1}\,{\Delta_0(s)} 
 		}}\spazio{1.2}\cr
 &{}& \phantom{{\it c0} \left( s \right) =}-\frac12\,{\frac { \sqrt{j}\,u_{{3}} \left( s \right) }{ 
 		\sqrt{2\,j+1}}}\mbox{}-\frac12\,{\frac { \sqrt{j+1}\,u_{{4}} \left( s \right) }{ 
 		\sqrt{2\,j+1}}}\spazio{1.2}\cr
&{}& c_1(s) =-{\frac { \sqrt{\Omega} \left( j+1 \right)  \sqrt{j}\,u_{{1}} 
		\left( s \right) }{ \sqrt{2\,j+1}\,{\Delta_2(s)} \mbox{}}}-{\frac { \sqrt{j+1} 
		\sqrt{\Omega}j\,u_{{2}} \left( s \right) }{\sqrt{2\,j+1}\,{\Delta_0(s)} 
		}}\spazio{1.2}\cr
&{}& \phantom{{\it c0} \left( s \right) =}-\frac12\,{\frac { \sqrt{j}\,u_{{3}} \left( s 
		\right) }{ \sqrt{2\,j+1}}}\mbox{}+\frac12\,{\frac { \sqrt{j+1}\,u_{{4}} \left( s \right) }{ 
		\sqrt{2\,j+1}}}\spazio{1.2}\cr
&{}& d_0(s) =-{\frac { \sqrt{j+1} \sqrt{\Omega}j\,u_{{1}} \left( s \right) }{ 
		\sqrt{2\,j+1}{\Delta_2(s)}}}\mbox{}-{\frac { \sqrt{j} \left( j+1 \right)  
		\sqrt{\Omega}\,u_{{2}} \left( s \right) }{\sqrt{2\,j+1}\,{\Delta_0(s)} }}\spazio{1.2}\cr
&{}& \phantom{{\it c0} \left( s \right) =}+\frac12\,{\frac { \sqrt{j+1}\,u_{{3}} \left( s 
	\right) }{ \sqrt{2\,j+1}}}\mbox{}-\frac12\,{\frac { \sqrt{j}\,u_{{4}} \left( s \right) }{ 
	\sqrt{2\,j+1}}}\spazio{1.2}\cr
&{}& d_1(s) =-{\frac { \sqrt{j+1} \sqrt{\Omega}j\,u_{{1}} \left( s \right) }{ 
		\sqrt{2\,j+1}\,{\Delta_2(s)} }}\mbox{}+{\frac { \sqrt{j} \left( j+1 \right)  
		\sqrt{\Omega}\,u_{{2}} \left( s \right) 
	}{\sqrt{2\,j+1}\, {\Delta_0(s)}}}\spazio{1.2}\cr
&{}& \phantom{{\it c0} \left( s \right) =}+\frac12\,{\frac { \sqrt{j+1}\,u_{{3}} \left( s 
	\right) }{ \sqrt{2\,j+1}}}\mbox{}+\frac12\,{\frac { \sqrt{j}\,u_{{4}} \left( s \right) }{ 
	\sqrt{2\,j+1}}}
\nonumber
\end{eqnarray}
$\blacktriangleright$ The relations for the ``odd'' states
\begin{eqnarray}
&{}& \displaystyle a_0(s) = a_1(s) =\frac12\,u_{{1}} \left( s 
\right)\spazio{1.2}\cr
&{}& \displaystyle b_0(s) =-b_1(s) =\frac12\,u_{{2}} \left( s 
\right)\spazio{1.2}\cr
&{}& c_0(s) =-{\frac { \sqrt{\Omega}\mbox{} \sqrt{j+1} \sqrt{j}}{ \sqrt{2\,j+1}} 
\biggl( {\frac { \sqrt{j+1}\,u_{{1}} 
			\left( s \right) }{{\Delta_2(s)} }}-{\frac { \sqrt{j}\,u_{{2}} \left( 
			s \right) }{{\Delta_0(s)}  }} \biggr) }
\spazio{1.2}
\cr
&{}& 
\phantom{ {\it c0} \left( s \right) =}-\frac12\,\frac {1}{ 
		\sqrt{2\,j+1}} \,\biggl[\, \sqrt{j}\, \biggl( 1+{\frac { \bigl( \Omega\,s+2 \bigr) 
		s}{{\Delta_0(s)}}} \biggr) \,u_{{3}} \left( s \right)  
\spazio{1.2}\cr
&{}& 	
\phantom{ {\it c0} \left( s \right) =}+ \biggl( 1+{\frac 
		{ \bigl( \Omega\,s+2 \bigr) s}{{\Delta_2(s)} }} \biggr) \,u_{{4}} 
	\left( s \right)  \sqrt{j+1}\mbox{} \,\biggr]
\spazio{1.2}
\cr
&{}& 
c_1(s) =-{\frac { \sqrt{\Omega}\mbox{} \sqrt{j+1} \sqrt{j}}{ \sqrt{2\,j+1}} 
\biggl( {\frac { \sqrt{j+1}\,u_{{1}} 
			\left( s \right) }{{\Delta_2(s)} }}+{\frac { \sqrt{j}\,u_{{2}} \left( 
			s \right) }{{\Delta_0(s)}  }} \biggr) }
\spazio{1.2}\cr
&{}&
\phantom{ {\it c0} \left( s \right) =}-\frac12\,\frac {1}{ 
		\sqrt{2\,j+1}}\, \biggl[\,  \sqrt{j} \left( 1-{\frac { ( \Omega\,s+2 ) 
			s}{{\Delta_0(s)} }} \right)\, u_{{3}} \left( s \right) 
\spazio{1.2}\cr
&{}&	
\phantom{ {\it c0} \left( s \right) =}	- \sqrt{j+1} 
	\biggl( 1-{\frac { ( \Omega\,s+2 ) s}{{\Delta_2(s)} }} 
	\biggr)\, u_{{4}} \left( s \right)\mbox{}\, \biggr]
\spazio{1.2}\cr
&{}&
 d_0(s) =-{\frac { \sqrt{\Omega}\mbox{} \sqrt{j+1} \sqrt{j}}{ \sqrt{2\,j+1}} 
 \left( {\frac {\sqrt{j}\,u_{{1}} \left( s 
 			\right)  }{{\Delta_2(s)}}}+{\frac {\sqrt{j+1}\,u_{{2}} \left( s\right)  }
 		{{\Delta_0(s)} }} \right) }
\spazio{1.2}\cr
&{}& 
\phantom{ {\it c0} \left( s \right) =} +\frac12\,\frac {1}{ 
 		\sqrt{2\,j+1}}\, \biggl[\,  \sqrt{j+1} \biggl( 1+{\frac { \left( \Omega\,s+2 \right) 
 			s}{{\Delta_0(s)}  }} \biggr)\, u_{{3}}\left( s \right) 
\spazio{1.2}\cr
&{}&  	
\phantom{ {\it c0} \left( s \right) =} 	 - \sqrt{j} 
 	\biggl( 1+{\frac { \left( \Omega\,s+2 \right) s}{{\Delta_2(s)} }} 
	\biggr) \,u_{{4}} \left( s \right) \mbox{} \,\biggr] 
\spazio{1.2}\cr
&{}& 
d_1(s) =-{\frac { \sqrt{\Omega}\mbox{} \sqrt{j+1} \sqrt{j}}{ \sqrt{2\,j+1}} 
	\biggl( {\frac {\sqrt{j}\,u_{{1}} \left( s 
			\right)  }{{\Delta_2(s)}  }}-{\frac {\sqrt{j+1}\,u_{{2}} \left( s 
			\right)  }{{\Delta_0(s)} }} \biggr) }
\spazio{1.2}\cr
&{}& 
\phantom{ {\it c0} \left( s \right) =} +\frac12\,\frac {1}{ 
		\sqrt{2\,j+1}}\, \biggl[ \, \sqrt{j+1} \left( 1-{\frac { \left( \Omega\,s+2 \right) 
			s}{{\Delta_0(s)} }} \right)\, u_{{3}} \left( s \right) 
\spazio{1.2}\cr
&{}&	
\phantom{ {\it c0} \left( s \right) =}	+ \sqrt{j} 
	\biggl( 1-{\frac { \left( \Omega\,s+2 \right) s}{{\Delta_2(s)}}} 
	\biggr)\, u_{{4}} \left( s \right) \mbox{}\, \biggl] 	
\nonumber
\end{eqnarray}

\bigskip\bigskip\bigskip


\begin{thebibliography}{999}
\bibitem{Nota2} Although it is of no concern here since we are dealing with equal masses of the two 
quarks, we take however the opportunity of correcting a trivial and evident misprint we did in  
paper \cite{GSM} in reporting the formulas we had used. The last term of the definition of 
$D_\varepsilon(s)$ in (5) reads $2\rho^2\,s/[sh(s)-2\varepsilon b]$.
\end{thebibliography}
\end{document}